# Large, light-induced capacitance enhancement in semiconductor junctions simulated by capacitor-resistor nets

B. Vainas

The Weizmann Institute of Science, Rehovot (Israel)

**Abstract**

The equivalent circuit simulation of random resistors–capacitors (R-C) net, modified to include large capacitors interfacing between the random R-C bulk and the electrode surface, shows an enhancement of 3 orders of magnitude of the apparent real dielectric constant at low frequencies upon an introduction of resistors' percolating paths in the bulk.
The appearance of the bulk R-percolating paths can represent the photo-generated high conductivity state of semiconductor's bulk, an effect supported by the experimental observation that, in parallel with the photo-enhancement of the real dielectric constant, its imaginary part is strongly enhanced as well. The addition of the photo-generated charge carriers strongly enhances bulks' electrical conductivity, effectively confining the space charge region to the interface between bulk's edge and the electrode.
That could be a simple phenomenological explanation for the apparent dielectric constant enhancement upon illumination in photocells [1], not involving elaborate physical models.

**Introduction**

An interesting example of a large photo-capacitive effect has recently been reported in the literature [1]. Authors use a solar cell configuration. The bulk semiconductors are organic-inorganic lead tri-halide perovskites, in between electrically conducting electrodes.
The first significant characteristic of the reported effect is that the real part of the apparent dielectric constant measured at different frequencies was shown to increase 3 orders of magnitude upon illumination at low frequencies of 0.1 - 1 Hz, and the imaginary part has a similar, parallel increase upon illumination, meaning that the dissipation-ohmic component was enhanced as well.
The second significant characteristic is the strong increase in the conductivity at frequencies higher than about 100 kHz, as shown in the log-log plot of conductivity vs. frequency in Fig. 3b, of reference [2]. It can be interpreted as a power law electrical response, which is typical to non-homogeneous materials [3,4].

The simulation suggested here does not pre-suppose any particular analytical method, or physical model, other than a random R-C net structure in-between the electrodes of the solar cell. It nevertheless succeeds in replicating the main results - the strong, 3 orders of magnitude enhancement of the real part of the apparent dielectric constant (capacitance) upon illumination, and the power-law conductivity characteristics at high frequencies.

**Experimental**

All simulations were made with, SIMetrix/SIMPLIS Intro, 2011, by measuring the AC current frequency response of a random resistor-capacitor net, that was used before [3,4]. In all the cases of recording the in-phase current / out of phase current / phase angle of the current relative to the driving ac voltage source, the current signal output was measured as a voltage drop across a small (1 milliohm) shunt resistor between the left hand side electrode (R1, Fig. 1) and ground.

Given that the present model aims at showing only the relative effect - the 3 order of magnitude enhancement of the dielectric constant upon illumination, the voltages measured on R1 are not converted to the corresponding absolute values of current and conductivity, but used directly, or divided by a frequency reference voltage, to get arbitrary unit capacitance, and then applied to vertical axes of log(arbitrary units value) vs. log(frequency) plots.

The frequency reference voltage by which the in-phase, or out of phase components of the current (expressed as voltages), are divided to obtain dielectric data, is the voltage drop on the small shunt resistor in an auxiliary, simple series C-R circuit (C24-R98, at the right side of Fig. 1), connected to the AC voltage source, in parallel with the net.

The real capacitance, C, is derived from the imaginary part of conductance, Sigma'',

Sigma'' = i$\omega$C, (eq. 1)

where $\omega$, denotes the angular frequency, and i, is the square root of -1. Therefore, given Sigma'', we can derive arbitrary units C, for use in log-log plots, by dividing the voltage proportional to Sigma'' by a voltage proportional to frequency.

Given that the macroscopic geometry of the solar cell (electrodes' area, etc.) can be reasonably assumed to be independent of illumination conditions, the real part of the dielectric constant will be used interchangeably with capacitance.

As will be suggested below, the enhancement of the capacitance can be the result of the diminished distance between charges and their countercharges, rather than the enhancement of the dielectric constant, based on some intricate physics.

Fig. 1

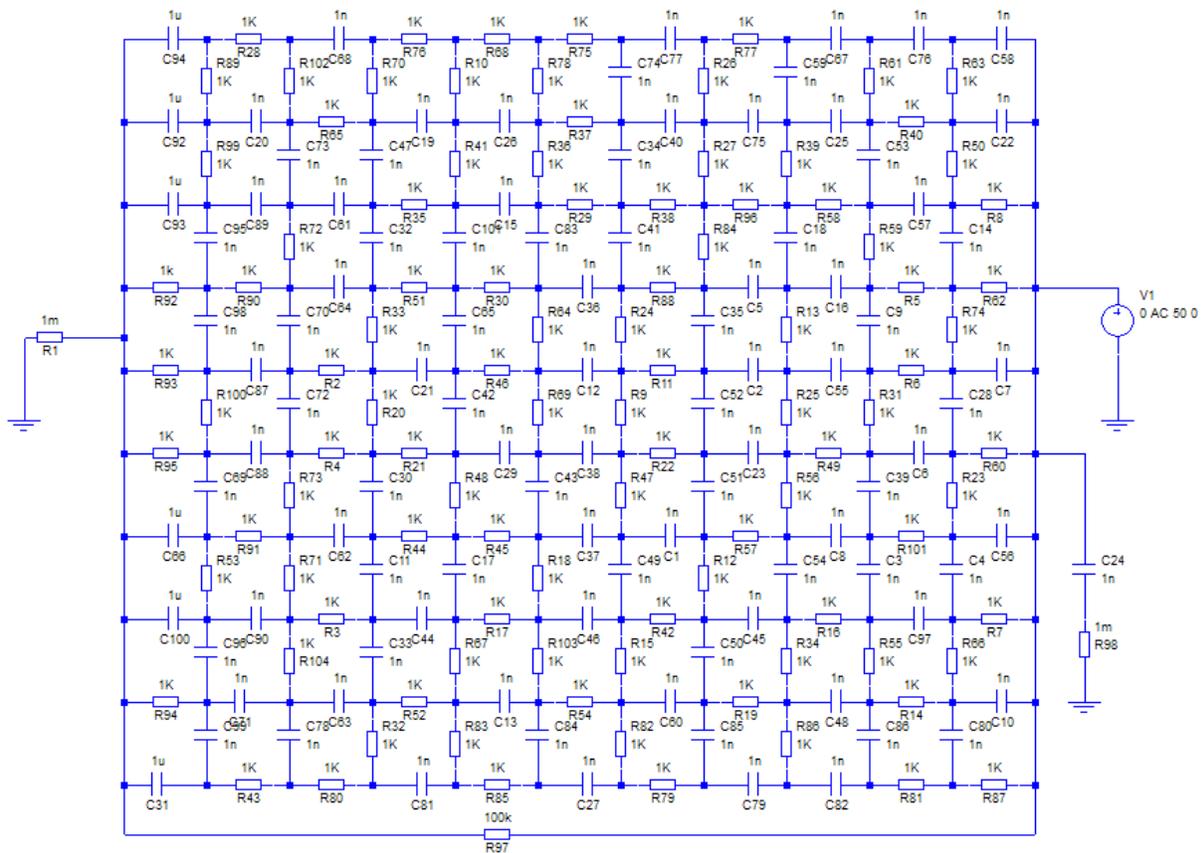

Fig. 1: Simetrix analog circuit simulator schematic. An AC analysis of a random R-C net, having an equal number of resistors and capacitors (100:100). All capacitors directly attached to the left hand side, ground electrode, are 1 microF, while all the rest, bulk capacitors, are 1 nanoF.
This particular net has two distinct continuous, all-resistors pathways, starting from R62 and terminating at several large capacitors on the left hand side electrode. There is a large "leakage" resistor (R97, 100 Kohm) inserted between the two electrodes.

## Results and Discussion

The distinctive feature of the photo-effect studied here is the simultaneous increase in conductivity accompanying the capacitance enhancement. As shown in [1] fig. 1b, and in fig. 3 of the present work, the imaginary part of the dielectric constant (the ohmic component), are enhanced 3 and 1 orders of magnitude, respectively.

Photoconductivity is a well established and widely observed effect, that involves photo-generation of mobile carriers by, for example, a sub-bandgap excitation of deep donors or acceptors. One can argue that carriers' density enhancement following illumination can be expected to diminish the width of the space charge on the semiconductor side of the junction. In the case, of a doped semiconductor material, in a Schottky barrier type of junction, at a constant, non-flat band level applied voltage, one can expect an enhancement of capacity, given by the Mott-Shottky expression [5]:

$$1/C^2 = (2/q*N*eps*eps_0)*(abs(V - V_{fb}) - kT/q), \qquad (eq.\ 2)$$

Where, $eps_0$ is the vacuum permittivity, V the applied voltage, $V_{fb}$ the flat band voltage, eps is the relative dielectric constant, q the elementary charge and N the concentration of free charge.

In the present model, the mechanism that can be applicable to the enhancement of C by illumination-enhanced conductivity (or, N increase as in eq. 2) is a reduced width of the space charge, as the result of the large, left terminal attached capacitors becoming dominant in the case of being connected in series with conductive pathways, that were formed by the enhancement of conductivity upon illumination.

A simple, 1-D, comparison can be made between the low and high conductivity cases:

a. C-r-c-r-c-r-c-r-c-r-c representing the large, left-terminal attached capacitor, C, in series with smaller capacitors in series with resistors in the bulk.

b. C-r-r-r-r-r-r-r-r-r representing the same large terminal capacitor terminating resistors percolating path.

The impedance in case a: at low frequencies, resistors' contribution to impedance will be negligible, resulting in an effective connection of a large C with much smaller equivalent capacitance of small capacitors connected in series. The result is a small overall capacitance.

The impedance in case b: the capacitance is clearly dominated by the large capacitance, C, which can be viewed as a capacitance enhancement in the context of the present model.

At high frequencies, the impedance in both cases becomes progressively dominated by resistors, over capacitors.

The 1-D comparison was verified using the simulation setup based on that given in Fig. 1, but replacing the 2-D net with 1-D chains. Fig. 2 clearly demonstrates the expected result: the case "b", the high conductivity analog, shows a more than 3 orders of magnitude enhancement of capacitance at lower frequencies.

Fig. 2

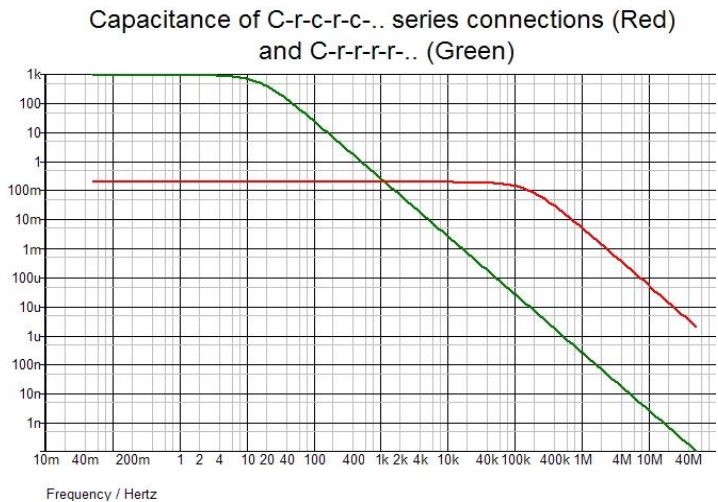

Fig. 2: Arbitrary units capacitance derived from the imaginary part of the current, measured as a voltage drop across a small shunt resistor, and divided by a voltage proportional to frequency. C – 1 microF, c – 1 nanoF, r – 1 Kohm. There are 5 small capacitors and 5 resistors in the first case (red trace), and 10 resistors in the second case (green trace). In both cases there is a single large capacitor, C.

The results of the simple 1-D model shown in Fig. 2 are not significantly different, in the context of low frequency capacitance (dielectric constant) enhancement, from the result for the 2-D random R-C net, in Fig. 3 below. In both these cases we observe a ~3 orders of magnitude enhancement of capacitance at frequencies lower than about 100 Hz.

The simulation in Fig. 3 was first made (green line) with the net containing the two R-percolation paths, as they appear on Fig. 1 above, and then "cutting" these paths (red line) by replacing a resistor R62 with a capacitor.

The 0.5 downward slope, in Fig. 3, from about 10 KHz, to higher frequencies is a clear result of the 0.5 power law characterizing the impedance of 2-D random R-C networks of the 1:1 R:C composition ratio [3,4].

Fig. 3

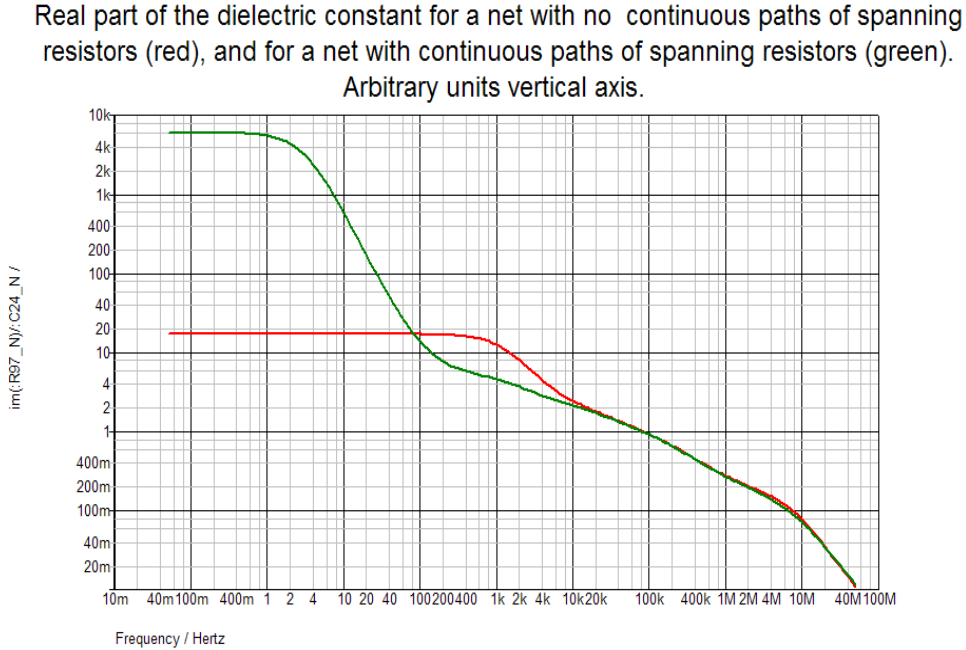

Real part of the dielectric constant for a net with no continuous paths of spanning resistors (red), and for a net with continuous paths of spanning resistors (green). Arbitrary units vertical axis.

The real part of AC conductivity as function of frequency was reported in Fig. 3b of [2], based on the original work on the solar cells [1]. There is a striking similarity between the simulation result given in Fig. 4 below, and the result reported in [2]. This similarity can suggest that the nearly flat, raised part of conductivity (illumination, green line) in Fig. 4, and the corresponding nearly flat region in Fig. 3b of [2] can be explained by the conductivity percolation pathways. Conductivity percolation can characterize the response at frequencies lower than that, indicated by the lower frequency edge, at around 10 KHz, of the power law region, as can be noticed in Fig. 4. Above this frequency, we have a 0.5 power law, as expected [3, 4] from the 100:100 R:C composition of the random R-C network in Fig. 1.

It should be noted that the upward turn of conductivity at high frequency in Fig. 3b of [2] has only a qualitative resemblance to the clear power law at high frequencies in Fig. 4. Extending the experimental frequency range towards higher frequencies could, in principle, validate the power law suggestion, and the random R-C model associated with it.

Fig. 4

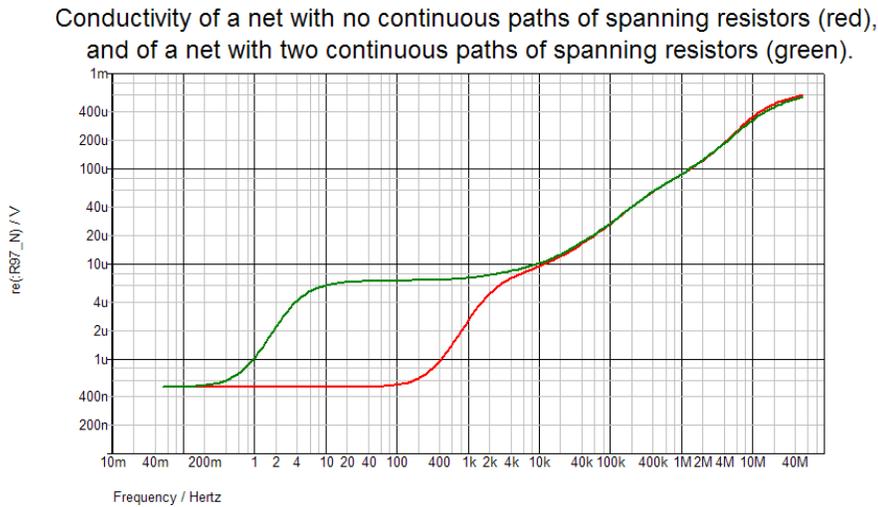

At a given frequency, the conductivity at low frequencies, around 100 Hz in Fig. 4, is increased by about an order of magnitude by the transition to the "high conductivity" state created by the resistors' percolation paths in the bulk. The imaginary dielectric constant frequency response can then be expected to show two parallel lines, inclined with slope -1, and the enhancement, expressed by the relative vertical position of the lines at a given frequency around 10 to 100 Hz would then be about 1 order of magnitude. That has indeed been observed in simulations (not shown here). So, the enhancement of the imaginary part of the dielectric constant is about 1 order of magnitude in present simulations, in comparison to ~3 orders of magnitude enhancement of the real part of the dielectric constant. It should be noted that no effort to optimize the present model has been made. Thus, only two resistors' bulk percolation paths are present in the "high conductivity" case (photocell under illumination).

## Summary

The series macroscopic connectivity, inherent to the M-S-M (metal-semiconductor-metal) type photo-diodes (like Schottky diodes), with front and back conductive electrodes, and a light absorbing semiconductor material in-between, suggest that the macroscopic interfaces between the phases, apart from creating the electrical field oriented to separate the photogenerated electron-hole pairs, could also drastically change the overall electrical characteristics of photodiodes as the result of switching-like effect of the change of conductivity upon illumination. This "photo-switching" can be considered as the first order effect that can explain the strong enhancement of the real part of the apparent dielectric constant, in particular, given the parallel enhancement of its imaginary part, which is equivalent to an enhanced conductivity, as been demonstrated in this work.